\documentclass[aps,prl,twocolumn]{revtex4}
\usepackage{graphicx}
\usepackage{bm}
\begin{document}
\def \tr{{\mbox{tr~}}}
\def \ra{{\rightarrow}}
\def \ua{{\uparrow}}
\def \da{{\downarrow}}
\def \be{\begin{equation}}
\def \ee{\end{equation}}
\def \ba{\begin{array}}
\def \ea{\end{array}}
\def \bea{\begin{eqnarray}}
\def \eea{\end{eqnarray}}
\def \nn{\nonumber}
\def \l{\left}
\def \r{\right}
\def \half{{1\over 2}}
\def \etal{{\it {et al}}}
\def \cH{{\cal{H}}}
\def \cM{{\cal{M}}}
\def \cN{{\cal{N}}}
\def \cQ{{\cal Q}}
\def \cI{{\cal I}}
\def \cV{{\cal V}}
\def \cG{{\cal G}}
\def \cF{{\cal F}}
\def \cZ{{\cal Z}}
\def \bS{{\bf S}}
\def \bI{{\bf I}}
\def \bL{{\bf L}}
\def \bG{{\bf G}}
\def \bQ{{\bf Q}}
\def \bK{{\bf K}}
\def \bR{{\bf R}}
\def \br{{\bf r}}
\def \bu{{\bf u}}
\def \bq{{\bf q}}
\def \bk{{\bf k}}
\def \bz{{\bf z}}
\def \bx{{\bf x}}
\def \bpsi{{\bar{\psi}}}
\def \tJ{{\tilde{J}}}
\def \W{{\Omega}}
\def \e{{\epsilon}}
\def \lam{{\lambda}}
\def \L{{\Lambda}}
\def \a{{\alpha}}
\def \t{{\theta}}
\def \b{{\beta}}
\def \g{{\gamma}}
\def \D{{\Delta}}
\def \d{{\delta}}
\def \w{{\omega}}
\def \s{{\sigma}}
\def \f{{\varphi}}
\def \x{{\chi}}
\def \e{{\epsilon}}
\def \h{{\eta}}
\def \G{{\Gamma}}
\def \z{{\zeta}}
\def \hatt{{\hat{\t}}}
\def \hn{{\bar{n}}}
\def \vk{{\bf{k}}}
\def \vq{{\bf{q}}}
\def \gk{{\g_{\vk}}}
\def \nd{{^{\vphantom{\dagger}}}}
\def \yd{^\dagger}
\def \av#1{{\langle#1\rangle}}
\def \ket#1{{\,|\,#1\,\rangle\,}}
\def \bra#1{{\,\langle\,#1\,|\,}}
\def \braket#1#2{{\,\langle\,#1\,|\,#2\,\rangle\,}}

\title{Dynamic projection on Feshbach molecules: a probe of pairing and phase fluctuations}
\author{Ehud Altman$^{1}$ and Ashvin Vishwanath$^{2}$}
\address{$^{1}$Physics Department, Harvard University, Cambridge, MA 02138.\\
$^{2}$Department of
Physics, University of California, Berkeley, CA 94720;}

\begin{abstract}
We describe and justify a simple model for the dynamics associated with
rapid sweeps across a Feshbach resonance, from the atomic to the molecular
side, in an ultra cold Fermi system. The model allows us to relate
the observed molecule momentum distribution, including its dependence
on the sweep rate, to equilibrium properties of the initial state.
For initial state near resonance, we find that phase fluctuations
sharply reduce the observed condensate fraction. Moreover, for very fast
sweeps and low temperatures, we predict a surprising nonmonotonic dependence
of the molecule condensate fraction on detuning, that is a direct signature of
quantum phase fluctuations.
The dependence of the total molecule number on sweep rate is
found to be a sensitive probe of pairing in the initial state, whether condensed
or not. Hence it can be utilized to  establish the
presence of a phase fluctuation induced `pseudogap' phase in these
systems.
\end{abstract}
\maketitle

Experiments with ultra cold fermions near a Feshbach resonance (FR),
opened a new window to the study of strongly interacting
condensates\cite{fesh_exp,JILA,MIT}.
They can access the strongly coupled
regime intermediate between weak pairing BCS superfluidity
and a BEC of molecules. In this region,
deviations from mean field theory {\em a la BCS}
\cite{eagles,leggett,holland},
are expected to be large. In addition, the ability to rapidly
vary the interaction parameters, provides a unique opportunity to study
quantum dynamics far
from equilibrium\cite{levitov,leo,galitzki}.
The focus of recent experiments has been to utilize dynamics,
in this case rapid magnetic field sweeps across the FR,
in order to probe the {\em equilibrium} properties of the
condensate in the crossover
region\cite{JILA,MIT}.
The idea was to convert cooper pairs in the initial state,
which would otherwise unbind during free expansion, into molecules.
The fact that the molecule momentum distribution depended on the
start position has been offered as strong indication that
indeed properties of the initial equilibrium state were being accessed.
However, precise connections to such properties are lacking.
\begin{figure}[t]
\includegraphics[width=7cm]{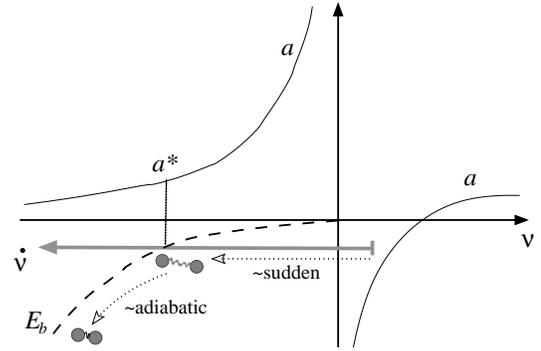}
\caption{{\em Model dynamics.} Field sweep is
effectively sudden up to detuning $\nu_\star$
depending on the sweep rate
$\dot\nu$. At this point the scattering length (and molecule size)
is $a_\star$. The molecule is assumed to evolve adiabatically from there.}
\label{fig:scheme}
\end{figure}

In this letter we formulate an approach that allows us to relate
equilibrium properties of such strongly interacting fermions to the
measurements in the dynamical experiments. Our approach
targets the regime of 'fast' magnetic field sweeps. Roughly, these are
ramp rates at which
the conversion efficiency of atoms into
molecules changes significantly with changing ramp rate.
Empirically
in the $^{40}$K system this implies a ramp rate 
faster than $1/20 \;G/\mu sec$\cite{JILA},
while it was too fast to be accessed in $^{6}$Li\cite{MIT}. The physics in this
regime turns out to have a remarkable simplicity, and for this
reason we will focus on it in this letter, although the bulk
of experiments to date have been done at slower ramp speeds.
It is hoped that the results of the present study will stimulate
experimental investigation of this regime.

Consider for a moment the extreme limit of an infinitely fast sweep.
Then, we can apply
the sudden approximation and simply use the initial state to evaluate the
final population of molecules. This intuitive picture was advocated by
Regal \etal \cite{JILA} (see also ~\cite{stoof}). 
Using the same assumption, Diener and Ho\cite{ho} 
estimated the
fraction of condensed molecules
as a function of the detuning of the initial state from resonance.
The molecule momentum distribution in this approach is given by
\bea
n_m(\bq)=\av{b\yd_\bq b\nd_\bq}_i ~~~
b\yd_\bq=\int d\bk \f_f(k)c\yd_{{\bq\over 2}+\bk\ua}c\yd_{{\bq\over 2}-\bk\da},
\label{proj}
\eea
where $\f_f(k)$ is the molecular WF at the end of the ramp,
deep in the molecular side of the resonance. The average $\av{~}_i$
is taken over the initial state, which in Ref. \cite{ho}
was assumed to be a BCS WF, calculated within
the mean field theory of \cite{leggett}.
Indeed, this gives a bimodal momentum distribution,
with a peak at $\bq=0$ due to
cooper pairs projected to molecules; as well as molecules at $\bq\ne 0$
due to pairwise projection of atoms belonging to different cooper
pairs. Although this is roughly what is seen in the experiments, it is
unsatisfactory in two important respects. (i) Mean field
theory is used to calculate the initial state, whereas it is well
known that in the crossover regime near the resonance,
quantum and thermally induced phase fluctuations play a very important
role. (ii) The approach ignores the dynamical aspects
of the experiment and is by definition unable to predict the
dependence of measured quantities (eg. conversion efficiency and
molecular condensate fraction) on ramp rate.

Both these issues are addressed in this paper.
The first, by
going
beyond the mean field theory, using the RPA \cite{randeria} to include
Gaussian phase fluctuations (or ``non condensed cooper pairs'').
In the BCS to BEC crossover regime these sharply reduce
the observed condensate fraction relative to the mean
field result of Ref. ~\cite{ho},
even at zero temperature. For fast sweeps at low temperature
this leads us to predict a surprising non monotonic
behavior of the molecule condensate fraction
versus detuning of the initial state from resonance.
Point (ii) raised above is addressed by an effective
model for the dynamics.
For fast sweeps we argue that the
time evolution of the system can be approximated by a `sudden' part
followed by an `adiabatic' time evolution part. The point at which the
time evolution changes character depends on the ramp rate,
the sudden evolution
persisting for longer at higher ramp rates.
Thus, the dynamics is approximated by projecting the initial
state of the system onto an {\em effective} molecular WF,
which is determined by the ramp rate. So, we will be using
Eq. (\ref{proj}), but with the effective molecular WF, $\f$
that depends on the ramp rate.
In this way we are able to
obtain the parametric dependence of various measured
quantities, such as conversion efficiency into molecules and molecular
condensate fraction  on the ramp rate,
and the parameters of the initial state.
At these fast sweep rates,
we find that conversion of Cooper pairs into molecules
is vastly more efficient than that of uncorrelated
pairs of atoms. This applies both to condensed and
uncondensed Cooper pairs, and ultimately
results from a short distance singularity present in the
Cooper pair wavefunction (WF), that allows them to have a
non-negligible overlap even with small sized molecules.
Since these dynamical measurements are equally
sensitive to noncondensed Cooper pairs, they can potentially
probe the phase fluctuation induced pseudogap
phase (where thermal and quantum phase fluctuations
destroy the condensate but pairing remains). Details such as the
momentum distribution of noncondensed pairs may also be accessed.

We now motivate the model dynamics via physical arguments and derive the
consequences for the system of interest. Later we perform a
nontrivial check by
showing that this simple scheme indeed reproduces the physics in a nontrivial
toy model (the Dicke model) which is
solved numerically without approximations.
In what follows we concentrate on a {\em wide}
FR, relevant to experiments in $^{40}K$\cite{JILA}
and $^7Li$\cite{MIT,innsbruck}. That is,
$g_s\equiv g\sqrt{n/2}>>\e_f$  where $g$ is the
coupling between the open and closed channels and $n$
is the atom density.
For most purposes it
is then possible to neglect the
occupation of closed channel molecules. The problem reduces
to spin-$\half$ fermions interacting via a
contact potential
\be
H = \int dx c^{\dagger}_\sigma (x)
(-\frac{\nabla^2}{2m}-\mu)c_\sigma(x) - u c^{\dagger}_\uparrow(x)
c^{\dagger}_\downarrow(x) c_\downarrow(x) c_\uparrow(x)
\label{Hamiltonian}
\ee
Ultraviolet divergences are avoided in the standard fashion
by writing all physical results
in terms of an s-wave scattering length $a$ instead of $u$.
Near resonance the scattering length diverges as
$a \approx -mg^2/4\pi\nu$ where $m$ is the atomic mass,
and $\nu$ is the detuning
in energy units, which is related to the magnetic detuning
via the magnetic moment
difference $\Delta \mu$ between closed and open channels:
$\nu=\D\mu (B-B_0)$.

Our two stage approximation of the dynamics is depicted graphically in
Fig. \ref{fig:scheme}. For slow sweeps,
atoms are converted adiabatically to weakly bound Feshbach molecules (FM)
\cite{regal}.
The binding energy of these molecules is $E_b=-1/(m a^2)\propto \nu^2$ 
(for $k_f a<<1$).
Now consider a change
in the detuning parameter at a constant rate $\dot{\nu}$.
Once the binding energy is large enough, such that $\dot{E}_b\ll E_b^2$,
the time evolution is expected to be adiabatic.
So, for a given sweep rate $\dot\nu$,
there is a characteristic detuning $\nu_\star$,
which marks a dynamic crossover  for the system.
At $\nu_\star$, $\dot{E}_b\approx E_b^2$ ~\cite{long_paper}.
Our simplified two stage model for the dynamics approximates
the change up to $\nu_\star$ as sudden, while the rest
is considered as perfectly adiabatic.
Thus the WF
at the initial state is effectively
``projected'' on FMs that occur at detuning
$\nu_\star$. These molecules evolve adiabatically into more tightly bound
ones corresponding
to the final value of the field, while the rest of the atomic
population remains unbound.

Within this model, a faster sweep rate simply moves
the break point $\nu_\star$ to larger negative detuning,
leading to effective projection of the initial WF
on smaller FMs.
We can establish a precise connection between the sweep rate 
and the size of the
effective molecule WF.
The typical size of the FM at $\nu_\star$ is the scattering length $a_\star(\nu_\star)$.
Using $E_b=-1/(ma^2)$,
and the relation between the scattering length and the detuning 
$\nu$ quoted earlier,
with the break condition $E_b^2(a_\star) = dE_b(a_\star)/dt$ we obtain

\be
k_f a_\star=\left(\frac{3\pi}{4}\frac{g_s^2}{\dot\nu}\right)^{1/3}.
\label{a_star}
\ee

We now employ the two stage model
and Eq. (\ref{a_star}) to determine the main features of the
final molecule distribution and its dependence on both
initial state and sweep rate.
We need to evaluate the following correlation function,
in the initial state:
\be
n_m(\bq)=
\int d^3 k d^3 k' \f^*(k)\f(k')\av{c\yd_{{\bq\over 2}+\bk\ua}c\yd_{{\bq\over 2}-\bk\da}
c\nd_{{\bq\over 2}-\bk'\da}c\nd_{{\bq\over 2}+\bk'\ua}}.
\label{proj2}
\ee
The size of the molecular pair WF $\f(k)$ is $a_\star$,
as prescribed by the model dynamics.
First consider a mean field (BCS)
approximation of the initial state.
In this case the correlation function appearing in (\ref{proj2})
may be factorized to obtain, as in Ref. \cite{ho}:
\bea
n_m(\bq)&=&\left|\int d^3 k\f(k)\av{c\yd_{\bk\ua}c\yd_{-\bk\da}}
\right|^2 \d(\bq)\nn\\
&&+\int d^3 k|\f(k)|^2\av{n_{{\bq\over 2}+\bk\ua}}\av{n_{{\bq\over 2}-\bk\da}}
\label{nBCS}
\eea
The first contribution, proportional to the anomalous
expectation value,
gives the condensed part of the distribution.
In fact this is just the square of the overlap between
the final molecule WF and the ``cooper pair WF''.
The second, non condensed part, contains the normal expectation values.
To make further progress
we note that if the molecule size $a_\star$ is much smaller than
the inter-particle spacing, we can replace the exact WF
with a box WF in momentum space, of the same spatial extent.
We take
$\f(\bk)=\sqrt{3/4\pi}a_\star^{3/2}$ for $k<1/a_\star$
and $\f(\bk)=0$ outside this sphere. Now $\f(\bk)$
serves as a cutoff
to the relative momentum integrals.
Using this in (\ref{nBCS}), the number of normal molecules is found to be
\be
N_n=(N/2)(k_f a_\star)^3
\label{N_n}
\ee
where $N$ is the total atom number.
This result is easily understood
in terms of the overlap of a random pair of atoms in the Fermi sea, 
with a molecular WF. The conversion
efficiency is then proportional to the ratio of the
molecule volume to that
occupied on average by nearby atoms.
In contrast, the number of condensed molecules calculated from
(\ref{nBCS}) is
\be
N_0={6V a_\star^3 \over (2\pi)^2}\left|\int_0^{a_\star^{-1}}
 \frac{dk k^2 \D/2}{\sqrt{\D^2+
\xi_k^2}}\right|^2=\frac{9N}{8}\left({\D\over \e_f}
\right)^2 k_f a_\star
\label{N0bcs}
\ee
The last step relies on the high momentum divergence
of the integral, due to which it depends crucially on the high momentum cutoff.
The result (\ref{N0bcs}) will have important consequences
on interpretation of experiments, and it is worthwhile to remark
on its physical origin.
The number of condensed molecules is proportional to the square of the
overlap of the molecular WF $\f_m$ with the Cooper pair WF $\f_c$.
Now, the molecule WF is appreciable only within a region $r< a_\star$,
where $\f_m(r)\sim a_\star^{-3/2}$.
Since this region is smaller than the average inter-particle distance,
the molecules probe the Cooper pairs at
very short distances.  Now, the Cooper pair WF has a singular short
distance behavior, 
$\f_c(r)\propto 1/r$.
Therefore $|\langle \f_m|\f_c\rangle|^2\propto a_\star$
which is a much larger overlap than might be naively expected.
Note, in solid state systems where a natural short distance cutoff exists,
these features are absent and hence have not been emphasized.

Combining the mean field results for $N_0$ and $N_n$ we can
evaluate the condensate fraction in the molecule distribution:
\be
f_{MF}=\frac{N_0}{N_0+N_m}=\frac{1}{1+4/9(\e_f/\D)^2(k_f a_\star)^2}
\label{f_MF}
\ee
The dependence on the initial state enters this expression through the
factor $\e_f/\D$. For large positive detuning we expect weak pairing
$\D/\e_f\sim e^{-1/k_f a}$, while
close to resonance $\D\sim \e_f$.
For sufficiently fast sweep rates, $k_f a_\star<<1$
we have $f_{MF}\sim 1$ in the vicinity of the resonance. This is because
Cooper pairs are much more efficiently converted into molecules, 
and within the mean field approximation used above, 
all Cooper pairs are condensed.
However, especially in this region close to the resonance we expect 
phase fluctuations (Cooper pairs at finite momenta) to be excited, 
which will lead to $f<1$.

To obtain such corrections to the mean field result (\ref{nBCS})
we calculate the correlation function (\ref{proj2}) within
the RPA approximation. The details of the calculation are left
to \cite{long_paper}, here we briefly outline the main steps and the results.
Following Ref. \cite{randeria},
we consider a path integral representation of the partition
function $\cal{Z}$ determined by the Hamiltonian \ref{Hamiltonian}.
The interaction term may be decoupled with a
Hubbard-Stratanovitch  pair field $\D(\bq,\w)$.
In order to extract the desired correlation functions, we introduce a source
term $J(\bq,\w)\int d^4 k \f(k)c\yd_{{\bQ\over 2}+\bK}\ua c\yd_{{\bQ\over 2}-\bK}\da+ h.c.$,
where we have used the four vector notation $\bQ=(\bq,\w)$.
At $T>0$ the integral over imaginary frequencies is
converted into the usual Matsubara sum.  Then, the desired molecular
distribution function is:
\be
n_m(\bq)={1\over \cal{Z}}\sum_{\w\w'}\frac{\d ^2 \cal{Z}}{\d J(\bq,\w)\d J^\star(\bq,\w')}
\ee
While the saddle point approximation gives us the BCS result, 
here we go beyond and expand $\D(\bq,\w)=\D_0\delta(\bQ)+\h(\bq,\w)$, 
where $\D_0$ is the saddle-point value of the gap. 
The RPA approximation involves integrating out the fermions and 
expanding the resulting effective action to quadratic order in the 
$\h$ and $J$ fields. The poles in the
$\h$ propagator give the collective mode spectrum.
Finally, integrating out the $\h$ fields gives us an effective 
action solely in terms of $J$ from which the molecular distribution 
function $n_m(\bq)_{MF}+\d n_m(\bq)$ ($\d n(\bq)$ is the contribution 
due to fluctuations) may be evaluated.
\begin{figure}[t]
\includegraphics[width=7cm]{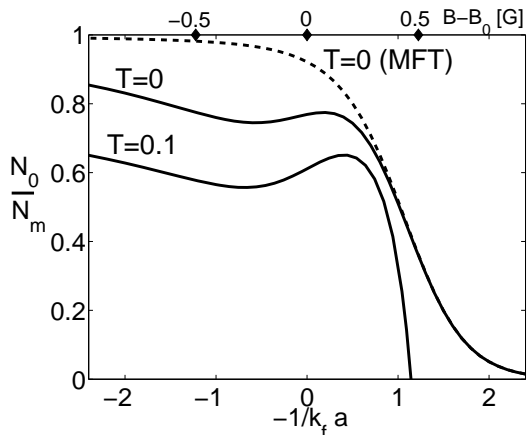}
\caption{Calculated molecule condensate fraction versus $-1/k_f a$ 
in the initial state (bottom axis), and versus magnetic 
field detuning for parameters of $^{40}$K~\cite{JILA} (top axis). 
The sweep rate corresponds to $k_f a_\star=0.3$.
Solid lines include the effect of quantum and thermal phase fluctuations. 
Dashed line shows the (T=0) mean field result for comparison.}
\label{fig:rpa}
\end{figure}


Assuming that the fluctuation contribution is dominantly 
from superfluid phonons, we can make a small momentum, 
small frequency expansion, where the $J$ propagator takes the form
$F(k_f a_\star, k_f a)[(c q)^2-(i\w)^2]^{-1}$, where $c$ is
the sound velocity in the superfluid and the function $F$
will be discussed shortly. It is important to note that the implied 
linear dispersion is only an approximation due to the low 
$q$ and $\w$ expansion.
At momenta $q\gtrsim 4mc$ the dispersion becomes quadratic. In addition
at $\w>\D$ phase fluctuations can decay into
quasiparticle excitations, leading to damping in the BCS limit
at $q\gtrsim \D/c\approx 1/\xi$

Carrying out the Matsubara
summation we obtain the RPA correction to the molecule
momentum distribution
\be
\d n(\bq)=F(k_f a_\star,k_f a)\frac{\coth (cq/2T)}{2cq}.
\label{dnq}
\ee
The number of non-condensed molecules due to these fluctuations
is found by integrating
over $\bq$ up to a natural cutoff. 
As discussed above, such a cutoff is provided
in the BCS limit by $q_1=1/\xi$, and in the BEC limit by $q_2=4mc$.
To cover the whole range we use the cutoff $q_0^{-1}=q_1^{-1}+q_2^{-1}$
which interpolates between the two limits.

The variation of the above fluctuation correction with sweep 
rate is encoded into
the dependence of the function $F(k_f a_\star, k_f a)$ 
on $k_f a_\star$. The leading dependence 
on  $k_f a_\star$ is found to be linear, which arises from the fact 
that non-condensed Cooper pairs are as efficiently transformed 
into molecules as condensed Cooper pairs.
By contrast, conversion of unpaired atoms into molecules is much 
less efficient and scales
as $(k_f a_\star)^3$.

The full function $F(k_f a_\star,k_f a)$ can be computed
along the crossover from BCS to BEC\cite{long_paper}.
Fig. \ref{fig:rpa} depicts the calculated condensed molecule fraction
including fluctuation corrections for sweep rate corresponding
to projection at $k_f a_\star=0.3$.
The nonmonotonic dependence on detuning toward the BCS limit
is due to competition between two effects.
On the one hand decreasing phase fluctuations act to
increase the condensed molecule fraction. On the other hand the ratio
$\e_f/\D$ appearing in the normal atom contribution diverges exponentially
at large detuning and eventually leads to vanishing of the molecular 
condensate fraction.

We now turn to a non trivial check of the two stage approximation
of the dynamics. 
Based on this model we found enhanced conversion
efficiency at rapid sweeps 
of cooper pairs (both condensed and non condensed) into FMs,
compared to unpaired atoms.
Using (\ref{a_star}), and the linear scaling in $k_f a_\star$
of the number of molecules arising from cooper pairs, we obtain
a parametric dependence on the sweep rate $\propto(g_s^2/\dot\nu)^{1/3}$. 
By contrast, from (\ref{N_n}) and (\ref{a_star}),
the conversion efficiency for unpaired atoms
is much lower, $\propto g_s^2/\dot\nu$.
We now wish to verify these non-trivial dependences using
a numerical simulation of the dynamics, not relying on the
two-stage model. We start from the 
microscopic two channel Hamiltonian\cite{holland}
\bea
H&=&\sum_{\bk\s}(\e_\bk-\mu)c\yd_{\s\bk}c\nd_{\s\bk}+
\sum_\bk(\e_k/2+\nu-2\mu)m\yd_k m\nd_k \nn\\
&&-g\sum_{\bk\bq}(m_\bq c\yd_{\ua\bk}c\yd_{\da \bq-\bk}+H.c.)+\mu N,
\label{H0}
\eea
which is regularized by absorbing the high momentum cutoff into renormalized
detuning parameter\cite{holland}.
Note, $m_\bq$ describes a (bare) closed channel molecule.

\begin{figure}[t]
\includegraphics[width=8cm,height=5cm]{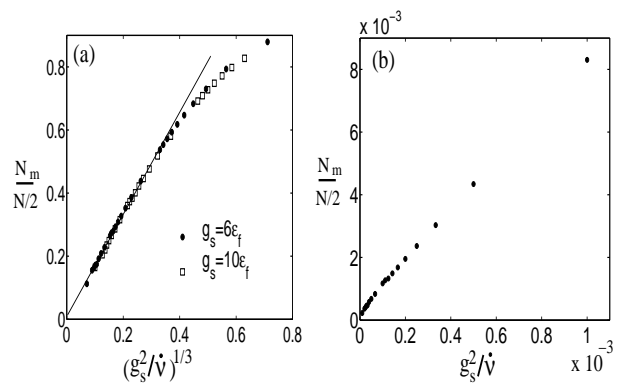}
\caption{Simulated mean Dicke model dynamics:
Final molecule number as a function of the sweep rate. 
(a) Initial state close to resonance (strong pairing).
The curves for different coupling constants 
$g_s$ fall on a universal curve $\propto 1/\dot\nu^{1/3}$ at fast rates,
in agreement with (\ref{a_star}) and (\ref{N0bcs}). 
(b) Initial state far detuned ($\nu=400\e_f$, $g_s=6\e_f$),
i.e. weak pairing. The  $1/\dot\nu$
behavior due to (\ref{N_n}) dominates.}
 \label{fig:dynamics}
\end{figure}

We compute the dynamics of (\ref{H0}) within the Dickie model
(i.e. keeping only $m_{\bq=0}$).  The initial state
is taken to be the equilibrium solution at detuning 
$\nu\ge 0$. 
Then $\nu$ is changed at a constant rate to far negative detuning where
the equilibrium population of $m_0$ would be $\sim 96\%$ and
the resulting dynamics (see e.g. \cite{leo}) 
is calculated numerically with no approximation. Following
the experiments we count the number of molecules in the {\em final} state,
which to a good approximation is $|\av{m_0}|^2$.
According to the two stage model this is directly related to 
the number of FMs produced after the sudden stage. 

The results are summarized
in Fig. \ref{fig:dynamics}.
Most importantly, for initial state at resonance (i.e. strong pairing),
the dependence on sweep rate fits $1/\dot\nu^{1/3}$, as expected
from our two stage model. By contrast, for weak pairing, 
deep in the BCS regime the
normal contribution $1/\dot\nu$ dominates (Fig. \ref{fig:dynamics}b).
The strong suppression of the prefactor of  $1/\dot\nu$
is an artifact of the Dickie model, which prohibits
occupation of molecules with $q\ne0$.

In summary, we presented a simple model of the dynamics of
rapid magnetic field sweeps across the FR. This allowed to relate 
the measurements of such dynamical experiments to properties 
of the initial equilibrium state of the Fermions.
For rapid sweeps at low temperatures in the crossover region 
we predict a
a non monotonic behavior of the final condensed molecule fraction
versus detuning, which is a direct signature of the quantum 
phase fluctuations in the initial state.
In addition,  the conversion efficiency to molecules at fast
sweep rates is argued to be a sensitive probe of pairing,
whether in a condensed state or not. This can be used to 
establish the presence of a fluctuation induced
pseudogap phase. In this regime we expect 
enhanced conversion efficiency due to the presence
of noncondensed fermion pairs, despite a vanishing condensate fraction. 

{\em Acknowledgements.} Discussions with E. Demler,
M. Greiner, D. Jin, W. Ketterle, D. Petrov, C. Regal, and  M. Zwierlein
are gratefully acknowledged. A.V. would like to thank the 
A. P. Sloan foundation for support.

\end{document}